\documentstyle[prb,aps,epsf,multicol]{revtex}

\begin{document}

\title{The fabrication of reproducible superconducting scanning tunneling microscope tips}

\author{O.~Naaman, W.~Teizer
and R. C.~Dynes\thanks{To whom correspondence should be
addressed.}}
\address{Department of Physics, University of
California, San Diego; 9500 Gilman Drive, La Jolla, CA 92093-0319}

\maketitle

\begin{abstract}
Superconducting scanning tunneling microscope tips have been
fabricated with a high degree of reproducibility.  The fabrication
process relies on sequential deposition of superconducting Pb and
a proximity-coupled Ag capping layer onto a Pt/Ir tip.  The tips
were characterized by tunneling into both normal-metal and
superconducting films.  The simplicity of the fabrication
process, along with the stability and reproducibility of the
tips, clear the way for tunneling studies with a
well-characterized, scannable superconducting electrode.
\end{abstract}
\begin{multicols}{2}
\section{Introduction}
Scanning tunneling microscopy (STM) has been proven to be an
important tool in the study of superconducting materials by
serving as a local probe of the density of states (DOS) near the
Fermi energy.\cite{pan00,yazdani97,renner95,truscott99} STM
studies of superconducting materials are conventionally performed
in the S/I/N (superconductor/insulator/normal-metal)
configuration, where a sharp normal-metal tip is brought to
within tunneling distance from a superconducting sample.  A
natural and intriguing extension of the capabilities of STM
studies would be to use a superconducting tip, thus allowing for
S/I/S measurements.

Although superconducting tips have been used in the past in an
attempt to form S-S point contacts,\cite{rodrigo94} true vacuum
tunneling has not been achieved until quite
recently.\cite{pan98}  The difficulty in fabricating
superconducting tips arises mainly from the conflicting needs of
making an atomically sharp tip, while keeping the tip apex clean
and well defined.  Since most known superconducting materials
readily oxidize, the fabrication of a superconducting tip is
significantly more complicated than that of a normal-metal tip,
where metals like Pt and Au can be used.  In addition, one may be
concerned that the tunneling current through a very small
junction area may lead to current densities that are sufficient
to destroy the superconducting state at the tip.  This issue was
addressed by Meservey who argued that currents, typical of normal
STM operation, could be carried without excessive depairing.
\cite{meservey88}

Pan {\it et al.} \cite{pan98} have demonstrated that quasiparticle
tunneling from a sharp superconducting tip is indeed possible.
They have used tips cut from Nb wire and cleaned {\it in-situ} by
field emission.  Although the work of Pan {\it et al.} provided an
important step towards the implementation of superconducting tips
in STM studies, their method of tip fabrication did not yield
sufficiently reproducible results, and a large variation in the
superconducting properties of their tips was
observed.\cite{pan98}  It may be argued that the field emission
cleaning procedure used did not remove all of the Nb oxides from
the tip.  There are a variety of oxides in Nb, out of which NbO$_
2$ and Nb$_ 2$O$_ 5$ are insulators and their removal can be
verified by successful tunneling at low temperatures. However,
NbO is metallic at 4.2 K and any residue of this material on the
tip apex, which is not directly detectable, may result in a
proximity layer of unknown thickness.  To our knowledge, no
method for the fabrication of reproducible superconducting tips
has been reported yet.

In this article we demonstrate a new, simple method for the
fabrication of superconducting tips.  By use of a controlled
Pb/Ag proximity bilayer deposited onto pre-cut Pt/Ir tips, we
have obtained superconducting tips with a high degree of
reproducibility. Furthermore, due to the slow oxidation rate of
Ag and the mechanical rigidity of Pt, our tips can be easily
manipulated {\it ex-situ} without any significant degradation of
their superconducting properties.

\section{Fabrication and characterization}
Pt$_{0.8}$/Ir$_{0.2}$ tips were cut from a 0.25 mm diameter wire.
We verified the quality of these tips by imaging the surface of a
highly ordered pyrolytic graphite single crystal at room
temperature.  The tips with the best resolution were selected,
and placed in a bell-jar evaporator such that the tip axis was
pointing in the directions of the evaporation sources to prevent
shadowing effects.  A 5000 {\AA} thick layer of Pb was deposited
on the Pt/Ir tips by thermal evaporation at a rate of $\sim$50
\AA/sec.  Without breaking vacuum, a thin layer of Ag was
subsequently evaporated at a rate of $\sim$2 \AA/sec.  The Ag
layer thickness was varied from 30-60 {\AA} in four different
deposition runs.  The pressure during the two-step evaporation
process was $\sim$4$\times$10$^{-6}$ torr or lower.

Pb (T$_c$=7.2 K) was chosen so that at 4.2 K the tips would be
well below their transition temperature and the layer thickness
was chosen to assure bulk superconductivity in the Pb layer
(${\xi_0}$=830 \AA).  The silver serves as a capping layer,
protecting the lead from rapid oxidation upon exposure to the
atmosphere.  The thin Ag layer is proximity-coupled to the lead
layer, and it has been

\begin{figure} \epsfxsize=\columnwidth
 \centerline{\epsfbox{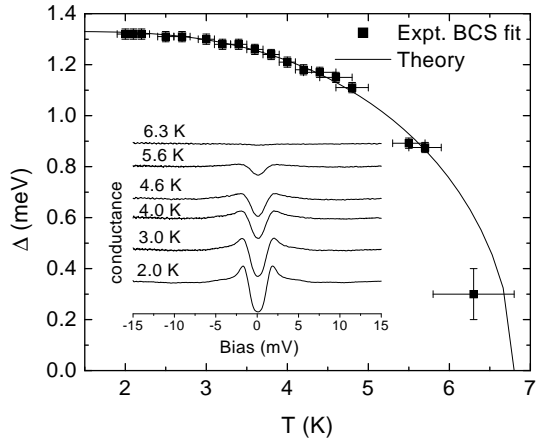}}
 \caption{Temperature dependence of the superconducting
gap for a tip with d$_{\rm{Ag}}$=30 {\AA}.  Inset- normalized
conductance curves at various temperatures.} \label{fig1}
\end{figure}

demonstrated\cite{katz95} that the Pb/Ag proximity junction is
reproducible and stable.  The Pb-Ag combination is also a good
metallurgical choice, as there is no significant alloying at the
interface.\cite{hansen58}  The tip is therefore expected to
behave as an effective superconductor with T$_c$ and $\Delta$
only slightly below that of bulk lead.

We have found that our tips can be stored in air for a period of
about two days before an oxide layer is formed, which is thick
enough to prohibit tunneling at low temperatures.  The lifetime
of the tips can be appreciably extended when stored in rough
vacuum, and we have successfully used tips after three weeks of
storage under these conditions.

In order to characterize the tips, we have performed tunneling
measurements on a Au sample in the temperature range 2.0-6.3 K,
using a custom built ultra high vacuum low temperature
STM.\cite{truscott99}  Ultra high purity $^4$He exchange gas was
used to thermally couple the STM to a liquid He bath, whose
temperature could be lowered to 2.0 K by lowering the He vapor
pressure.  Measurements above 4.2 K were performed while the STM
was allowed to slowly warm up.  Differential conductance was
measured using standard lock-in techniques.  At each bias voltage
sweep, 512 data points were taken at a rate of 3.0 ms per data
point.  Data was typically averaged over 250 (100) ramps below
(above) 4.2 K.  The junction resistance was 100 M$\Omega$ with
the bias voltage during the feedback cycle fixed at +50 mV with
respect to the grounded sample.  Since the Au sample is a normal
metal with a constant density of states (DOS) near the Fermi
energy, the obtained curves represent the tip DOS.

Typical results are presented in Figure\ \ref{fig1}.  Conductance
curves (shown in inset) were fitted to BCS theory using the Dynes,
Narayanamurti, and Garno DOS,\cite{dynes78} allowing both
$\Delta$ and $\Gamma$ to vary, while the temperature was fixed to
the value read by a thermometer mounted on the sample holder.  We
believe these values accurately reflect the

\begin{figure}
\epsfxsize=2.5in \epsfysize=3in
 \centerline{\epsfbox{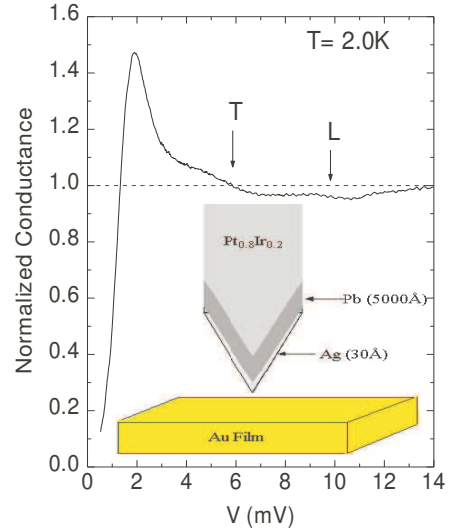}}
\vspace{0.5in}
 \caption{Conductance curve at 2.0 K for the same
tip as in Fig.\ \ref{fig1}.  The lead transverse (T) and
longitudinal (L) phonon features are clearly observed, their
position (V$_T$-$\Delta$=4.5 meV and V$_L$-$\Delta$=8.5 meV) and
magnitude consistent with literature (Ref. 13).  Inset- schematic
depiction of the experimental setup.} \label{fig2}
\end{figure}

temperature of the tip, as thermal coupling through the exchange
gas between the scanner head and the sample holder is good, and
self-heating is ruled out.\cite{heating}

A superconducting gap is clearly seen below 6.3 K, and its
evolution with temperature is very well described by the BCS
theory.\cite{tinkham96}  We obtain $\Delta_0$=1.33$\pm$0.01 meV
and T$_c$=6.80$\pm$0.05 K, yielding
${2\Delta_0/k_BT_c}$=4.54$\pm$0.05.  The values of $\Gamma$
estimated from the fits increased steadily with temperature, from
0.16 meV at 2 K to 0.40 meV at 6 K.  The values obtained for the
gap for tips from the same fabrication batch were in good
agreement with each other, while the variation of the gap value
among tips from different batches did not exceed 10\%, and was
consistent with the slightly different Ag layer thicknesses
produced in the separate depositions.  Below T=3.0 K it was
possible to resolve the lead phonon contribution to the tunneling
DOS, which arises from the energy dependent order parameter due
to strong electron-phonon coupling.  The phonon structure is
clearly seen around 2 K, and the obtained phonon energies and
structure (Fig.\ \ref{fig2}) are in good agreement with the data
in the literature.\cite{rowell69,arnold}  The fact that the
phonon structure is clearly observed, together with the
reproducibility of results for a range of junction resistances
and the characteristic exponential dependence of the tunneling
current on the tip-sample distance, is evidence that our tips form
single-step vacuum tunnel junctions.

\begin{figure}
\epsfxsize=\columnwidth
 \centerline{\epsfbox{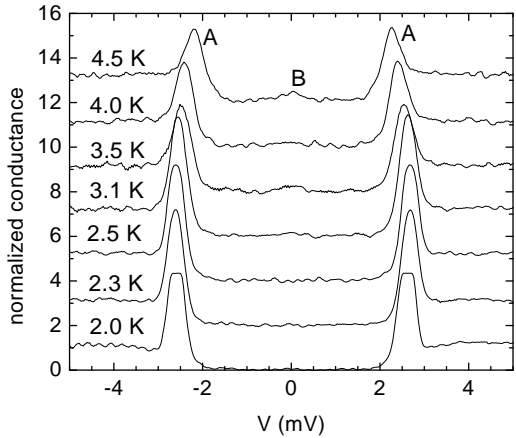}}
\caption{Normalized conductance at various temperatuers for an
S/I/S junction formed by a Pb(5000{\AA})/Ag(30{\AA}) tip over Pb
film (curves offset for clarity).  The coherence peaks at 2.0 K
are flattened by the saturation of our amplifier.} \label{fig3}
\end{figure}

Figure\ \ref{fig3} shows representative conductance spectra for
vacuum tunnel junctions formed between our superconducting tips
and a Pb film deposited {\it in-situ} on a graphite substrate.
These spectra are characteristic of an S/I/S junction, and
conductance peaks (A) are clearly visible at the expected voltages
corresponding to eV=$\pm$($\Delta_{tip}+\Delta_{Pb}$).  The peak
corresponding to the gap difference (B) disappears, as expected,
as the temperature is lowered and the number of available states
due to thermal excitations across the gap is reduced.  At higher
temperatures this peak is located at zero bias, which confirms
again that the superconducting gap value of the tip is very close
to that of lead.

It is remarkable that the tip structure is conserved despite the
relatively large amount of material deposited onto it.  Since our
STM is optimized for large area scans, and atomic resolution
images are therefore not ordinarily obtainable, we can only give
an estimate on the lower bound of the tip resolution.  This was
done by imaging the profile of step edges on graphite, and taking
the apparent width of the step to be characteristic of our
instrument resolution.  Using this procedure we obtain
resolutions of 5-20 {\AA}.  Since the growth of Pb is granular in
nature it is reasonable to assume that the tip resolution is
limited by the size of the Pb grain at the apex of the tip.
Limiting the mobility of Pb atoms during the deposition, by using
higher deposition rates, using a Ge seed layer, or by cooling
down the tip during deposition, would decrease the Pb grain size
and probably improve the tip resolution.

\section{conclusion}

In summary, we have demonstrated a new method for the fabrication
of superconducting tips.  We have shown that this method, relying
on proximity coupling between a Pb superconductor and Ag capping
layer, is capable of yielding reproducible, clean, and stable
tips with T$_c$ and $\Delta$ close to that of bulk Pb.  The
observed tunneling characteristics in both S/I/N and S/I/S
configurations and their temperature dependence are consistent
with expectations.

Many possible applications emerge from the availability of well
characterized, reproducible superconducting tips, ranging from
thermometry \cite{pan98} to spin polarized tunneling
\cite{meservey88}.  Recently there has been increased theoretical
interest in the use of superconducting tips for tunneling
experiments into high-T$_c$ superconductors\cite{zhao00,balatsky},
where for example, information regarding the order parameter
symmetry can be obtained by tunneling near a grain boundary in
such materials \cite{zhao00}.

Arguably, one of the most intriguing applications is the
possibility of forming a scannable Josephson junction
\cite{balatsky}.  Because of the inherently high resistance of a
vacuum tunnel junction, the Josephson coupling energy,
E${_J}\sim\hbar\Delta/e^{2}R_{NN}$, where $R_{NN}$ is the
junction normal resistance, is expected to be smaller than or
comparable to the thermal energy in the system. In this case,
strong phase fluctuations will have a dominant effect on pair
tunneling, as was shown by Ivanchenko and Zil'berman
\cite{ivan69}. Preliminary studies have shown \cite{naaman} that
low temperature I-V curves for junction resistances around 50-100
k$\Omega$ are in agreement with the theory of Ivanchenko
\textit{et al. }

To conclude, our newly developed tips may open the way to a
variety of STM studies previously out of reach, most notable is
the possibility for a scanning Josephson tunneling microscope.

\acknowledgements The authors would like to thank A. D. Truscott,
O. Bourgeois, and F. Hellman.  This work was supported by AFOSR
Grant No. F49620-92-J0070, ONR Grant No. N00014-92-J1320, and NSF
Grant No. DMR9705180.


\end{multicols}

\begin{references}
\bibitem{pan00} S. H. Pan, E. W. Hudson, K. M. Lang, H. Eisaki, S. Uchida, and J. C. Davis,
Nature (London) {\bf 403}, 746 (2000).

\bibitem{yazdani97} Ali Yazdani, B. A. Jones, C. P. Lutz, M. F. Crommie, D. M. Eigler,
Science {\bf 275}, 1767 (1997).

\bibitem{renner95} Ch. Renner and {\O}. Fischer, Phys. Rev. B.
{\bf 51}, 9208 (1995).

\bibitem{truscott99} A. D. Truscott, R. C. Dynes, L. F. Scheenmeyer, Phys. Rev. Lett. {\bf 83},
1014 (1999).

\bibitem{rodrigo94} see for example, J. G. Rodrigo, N. Agra\"{\i}t, S. Vieira, Phys.
Rev. B. {\bf 50} 374 (1994).

\bibitem{pan98} S. H. Pan, E. W. Hudson, and J. C. Davis, Appl. Phys. Lett. {\bf 73},
2992 (1998).

\bibitem{meservey88} R. Meservey, Physica Scripta
{\bf 38}, 272 (1988).

\bibitem{katz95} A. S. Katz, A. G. Sun, R. C. Dynes, K. Char, Appl. Phys. Lett. {\bf 66},
105 (1995).

\bibitem{hansen58} M. Hansen, {\it Constitution of Binary Alloys}, 2nd
ed.(McGraw-Hill, 1958)



\bibitem{dynes78} R. C. Dynes, V. Narayanamurti, and J. P. Garno, Phys. Rev. Lett. {\bf 41}, 1509 (1978).

\bibitem{heating} The power dissipated by quasiparticle relaxation
is inversely proportional to the junction resistance.  The
observed dI/dV curves for different junction resistances are
essentially identical, thus ruling out any significant
self-heating of the tip.  See also E. W. Hudson, Ph. D. Thesis,
University of California, Berkeley, 1999

\bibitem{tinkham96} M. Tinkham, {\it Introduction to Superconductivity}, 2nd
ed.(McGraw-Hill, New York, 1996)

\bibitem{rowell69} J. M. Rowell, in {\it Tunneling Phenomena in
Solids}, Ed. E. Burstein and S. Lundqvist (Plenum Press, New York,
1969)

\bibitem{arnold} Gerald B. Arnold, Phys. Rev. B. {\bf 18}, 1076
(1978).

\bibitem{zhao00} H. Zhao, and C.-R. Hu, Phys. Rev. B. {\bf 62}, 1308 (2000).

\bibitem{balatsky} J. \v{S}makov, I. Martin, and A. V. Balatsky,
arXiv:cond-mat/0009310 (2000).

\bibitem{ivan69} M. Ivanchenko and L. A. Zil'berman, Sov. Phys. JETP {\bf 28},
1272 (1969).

\bibitem{naaman} O. Naaman, W. Teizer, and R. C. Dynes, To be published

\end{references}
\end{document}